\documentclass[sn-mathphys-ay]{sn-jnl}

\usepackage{hyperref}
\hypersetup{breaklinks=true}
\usepackage{setspace}
\usepackage{fullpage}
\usepackage{indentfirst}
\usepackage{fix-cm}
\usepackage{graphicx}
\usepackage{fancyhdr}
\usepackage{amsmath}
\usepackage{amsfonts}
\usepackage{amssymb}
\usepackage[section]{placeins}
\usepackage{subcaption}
\usepackage{lineno}
\usepackage{rotating}
\usepackage{tabularx}
\usepackage{pdflscape}
\usepackage{afterpage}
\usepackage{capt-of}
\usepackage[table]{xcolor}
\usepackage{comment}

\newcommand{\bize}{\begin{itemize}}
\newcommand{\eize}{\end{itemize}}
\newcommand{\rct}{\color{black}}

\newcommand{\old}{\color{gray}}

\begin{document}

\title[{\rct The importance of exploration}]{{\rct The importance of exploration: Modelling site-constant foraging} 
}
\author[1]{\fnm{Sarah A.} \sur{MacQueen}}\email{macquees@gmail.com}
 
\author[2]{\fnm{Clara F.} \sur{Hardy}}\email{clarahardy@gmail.com}

\author[1]{\fnm{W. John} \sur{Braun}}\email{john.braun@ubc.ca}

\author*[1]{\fnm{Rebecca C.} \sur{Tyson}}\email{rebecca.tyson@ubc.ca}

\affil[1]{CMPS Department (Mathematics \& Statistics), University of British Columbia Okanagan, 3333 University Way, Kelowna B.C. V1V 1V7, Canada}
\affil[2]{University of Notre Dame, Sydney, Australia}

\abstract{\rct
Foraging site constancy, or repeated return to the same foraging location, is a foraging strategy used by many species to decrease uncertainty and risks.
It is often unclear, however, exactly how organisms identify the foraging site.
Here we are interested in the context where the actual harvesting of food is first preceded by a separate exploration period.
In this context, foraging consists of three distinct steps: (1) exploration of the landscape (site-generation), (2) selection of a foraging site (site-selection), and (3) exploitation (harvesting) through repeated trips between the foraging site and "home base".
This type of foraging has received scant attention in the modelling literature, leading to two main knowledge gaps.  First, there is very little known about {\it how} organisms implement steps (1) and (2).  Second, it is not known how the {\it choice} of implementation method affects the {\it outcomes} of step (3).  Typical outcomes include the foragers' rate of energy return, and the distribution of foragers on the landscape.
We investigate these two gaps, using an agent-based model with bumble bees as our model organism foraging in a patchy resource landscape of crop, wildflower, and empty cells.
We tested two different site-generation methods (random and circular foray exploration behaviour) and four different site-selection methods (random and optimizing based on distance from the nest, local wildflower density, or net rate of energy return) on a variety of outcomes from the site-constant harvesting step.
We find that site-selection method has a high impact on crop pollination services as well as the percent of crop resources collected, while site-generation method has a high impact on the percent of time spent harvesting and the total trip time. 
We also find that some of the patterns we identify can be used to infer how a given real organism is identifying a foraging site.
Our results underscore the importance of explicitly considering exploratory behaviour to better understand the ecological consequences of foraging dynamics.
}

\keywords{keyword1, Keyword2, Keyword3, Keyword4}
\keywords{Agent based model, bumble bee, crop pollination services, model mechanism, foraging site choice, insect foraging behaviour}

\maketitle

\section{Introduction}
\label{SectionCh3:Introduction}


 Repeated use of the same area, variously referred to as as site constancy, site fidelity, or path recursion, is a common behavioural strategy across many different organisms for a variety of purposes, but especially for foraging \citep{BergerTal2015}. Foraging site constancy has been observed in, for example, penguins \citep{Baylis2015}, sea lions \citep{Chilvers2008}, ants \citep{Crist1991}, and bees \citep{Thomson1997}. The benefits of site constancy for individuals include more efficient foraging, due to (i) memory of the site location and a greater familiarity with the spatial structure of resources within the site \citep{Osborne2001,Olsson2010}, (ii) increased mean reward from the patch  and reduced variance in the reward encountered during trips \citep{Possingham1989}, making trips both more profitable and more predictable, (iii)  better survivorship resulting from greater knowledge of the environment, which reduces predation risk \citep{Osborne2001,Brown2001}, and (iv) reduction in competition and conflict \citep{Paton1984,Garrison1999}. 
 
 There are many different factors that may affect how individuals choose from among potential foraging sites. These factors include distance from the central place \citep{Dramstad2003,Dramstad1996,Saville1997,Westphal2006}; net rate of energy return/intake \citep{Cartar1991,Cresswell2000}; foraging efficiency (ratio of energetic gain to expenditure) \citep{Charlton2010}; the type of resources available (preference for certain types, perhaps due to nutritional quality) \citep{Bobiwash2018,Ruedenauer2016}; resource density  \citep{Cartar2009,Dreisig1995}; and the presence of conspecifics \citep{Leadbeater2007}. In reality, individuals probably use some combination of the above factors, but for many organisms it is not known which factors dominate and under what conditions.

{\rct Modelling studies can be a powerful tool for elucidating how organisms use the landscape they inhabit \citep{patterson:2008, cobbold:2015, klappstein:2022, bracis:2015, mclane:2011, fagan:2022}.  Landscape-scale studies can be prohibitively expensive, and generating a sufficiently wide range of landscape configurations to test any particular theory can also be very difficult.  Models provide an ideal solution: A near-infinite range of landscape configurations can be investigated, and exact probabilistic foraging distributions or thousands of sample foraging paths can be generated.  The theory can then inform empirical work, allowing theory testing to be done with a much smaller array of landscapes. This approach, however, depends on there being a sufficiently detailed understanding of movement behaviour for model predictions to be useful.  Currently, it is generally not known exactly how any given organism identifies a foraging site, though some guidance is provided by optimal foraging theory \citep{pyke:1984, cezilly:1996, fortin:2002, courant:2012}.  It appears that memory plays a strong role in many cases \citep{ranc:2021, bracis:2015, kulakowska:2014}, and even insects have been shown to be capable of developing complex memories \citep{collett:2013}.
Modelling work has shown that, depending on the specific goals of the forager \citep{nauta:2020}, memory can be advantageous in certain contexts \citep{pull:2022, bracis:2015}, though is less beneficial than often thought \citep{arehart:2023}.  Existing models generally assume that the exploration and exploitation (harvesting) stages are concurrent \citep{fagan:2022, capera-aragones:2022, nauta:2020, avgar:2013, boyer:2010}, but many organisms, especially when they are naive, exhibit a separate exploration stage that is individually-driven \citep{larsen:1994, popp:2023, nowak:2023, osborne:2013, carter:2020, mandal:2019}.

The existence of separate exploration bouts means that the organism has the opportunity to sample many parts of the landscape to generate a mental list of possible foraging sites from which it will eventually select one.  
More specifically, in this context, site-constant foraging consists of three distinct steps: (1) exploration of the landscape (site-generation), (2) selection of a foraging site (site-selection), and (3) exploitation (harvesting) through repeated trips between the foraging site and "home base".
This type of foraging has received scant attention in the modelling literature, leading to two main knowledge gaps.  First, there is very little known about {\it how} organisms implement steps (1) and (2).  Second, it is not known how the {\it choice} of implementation method affects the {\it outcomes} of step (3).  Typical outcomes include the foragers' rate of energy return or fitness, and the distribution of foragers on the landscape.
Here we investigate two possible site-generation processes and four possible site-selection metrics, and observe how each pair changes the outcomes of site-constant foraging.  
Our goal is to determine when and how the processes (site-generation) and metrics (site-selection) involved in identification of a foraging site have a strong effect on these outcomes (e.g., fitness, distribution).
}
	
Bumble bees make an excellent case study for foraging site constancy. 
A high degree of site constancy in bumble bees has been well demonstrated in many field studies (e.g. \citet{Osborne2001,Cartar2004,Ogilvie2016,Pawlikowski2007,Saville1997}), and yet it is not currently known exactly how bumble bees choose their foraging sites. 	
Their behaviour results in pollen transfer between flowers of plants that require animal-mediated pollination, and is therefore very relevant to studies of crop pollination.  The pollination services provided by bumble bees can influence, e.g., fruit set, fruit size, and gene flow \citep{Osborne2001,Levin1979}.  Site constancy can be beneficial or detrimental to crops. The benefits include increased visits to the crop flowers, if the pollinators become site constant to the crop field, and increased pollen transfer to compatible stigmas \citep{Osborne2001}. Both of these effects lead to an increase in crop production (e.g. fruit set and berry weight for blueberries). The drawbacks include  decreased visits to the crop flowers, if the pollinators do not become constant to the crop field, reduced gene flow between patches \citep{Osborne2001}, and increased interspecific pollen transfer if the site contains multiple concurrently blooming species \citep{Ogilvie2016}.  Site constancy therefore has the potential to either increase or decrease crop pollination services, depending on the landscape and habitat conditions, and how these conditions affect bee foraging decisions. 

 Bumble bee foraging behaviour and its influence on crop pollination in general have been studied in a variety of observational and experimental field studies. There have been many field studies investigating the effect of wild flower enhancements on crop pollination (e.g. \citet{Feltham2015}, numerous studies reviewed by \citet{Haaland2011}), but these studies tend to focus on responses in terms of bee densities rather than directly measuring pollination services via flower visits or crop yield. Studies that measure crop yields as a result of pollinator behaviour are less frequent, but can be found (e.g. \citet{Benjamin2014,Button2014,Blaauw2014}).

Models have also been used to study bumble foraging behaviour and crop pollination services (e.g. \citet{Nicholson2019,Rands2014,Olsson2015}). Modelling studies enable prediction of pollination services, and allow for insights into the effects of bee behaviours and landscape differences on pollination services \citep{rouabah:2024}. They can also provide motivation to direct field studies. The results of a modelling study can point to significant traits, behaviours, or outcomes that can be measured in the field, and subsequently used to distinguish model alternatives, (e.g., \citet{Kendall1999}).  Similarly to the field studies, most modelling studies also focus on bee densities and abundances in crops, or visitation rates to different landscape areas (e.g. \citet{Lonsdorf2009,Becher2018,Haussler2017}), and do not explicitly model pollination service responses in terms flower visits or crop yields, though there are a few exceptions (e.g. \citet{Qu2018, Everaars2018, rouabah:2024}).

 Relatively few models of bumble bee foraging include site constancy. Those that do generally use a single metric for selecting foraging sites. For example \citet{Becher2018} and \citet{Lonsdorf2009} use distance from the nest, \citet{Qu2018} and \citet{Everaars2018} use variations on random searches, and \citet{MacQueen2021} uses resource preference. A study of how the metric used to select foraging sites affects the predictions of the model with respect to pollination services, behavioural budget, or fitness, has not, to our knowledge, been carried out.  
 In addition, when selecting a foraging site there must be generation of an initial set of options (``site list") from which the bee can choose.  We refer to the generation of the site list as the ``site-generation" method.
It has been observed that bees explore the landscape before settling on a foraging site \citep{Osborne2013,Woodgate2016}, and so it seems likely that each bee selects her foraging site from among the options found while exploring. This exploratory site-searching behaviour is often left out of foraging models, and the site list is instead generated using some more arbitrary method. It is therefore unknown if the measurements of pollination services reported by these models are broadly applicable, or specific to the chosen site-generation and site-selection methods. 

To investigate how site-generation and site-selection methods affect theoretical predictions of crop pollination and bee fitness, we develop a stochastic agent-based model (ABM) in which the various generation and selection methods are explicitly represented.  For parameterization purposes, we focus on bumble bees as our model forager.  We test our model on randomly-generated agricultural landscapes composed of crop, natural, and empty (no floral resource) areas.  For the crop areas, we assume the crop is highbush blueberry, a preferentially bumble bee-pollinated crop \citep{Javorek2002} that is often planted in large fields.  Natural areas in the landscape contain the floral resources of wild flowers.  We present an overview of our ABM in Section~\ref{SectionCh3:Methods}, and present the results and discussion in Sections~\ref{SectionCh3:Results} and~\ref{SectionCh3:Conclusion}.

\section{Methods}
\label{SectionCh3:Methods}

\noindent We use an expanded version of the  individual based model presented in \citep{MacQueen2021}, which simulates exploration and foraging trips by individual bumble bees, tracking pollination services.  In this earlier model, the searching pattern during exploration is a looping foray starting and ending at the nest, and a foraging site of constancy is selected from the sites identified during exploration based on maintaining a particular balance of the total amount of each resource type stored in the nest.  The expanded model presented here allows for different methods of site-generation and site-selection. We provide an overview below, highlighting the additions to the model, and refer the reader to the Supporting Information for a detailed description following the ODD format \citep{Grimm2006,Grimm2010}.

\subsection{Model Overview}
\label{SectionCh3:Model}

\begin{figure}
	\includegraphics[width=.5\textwidth]{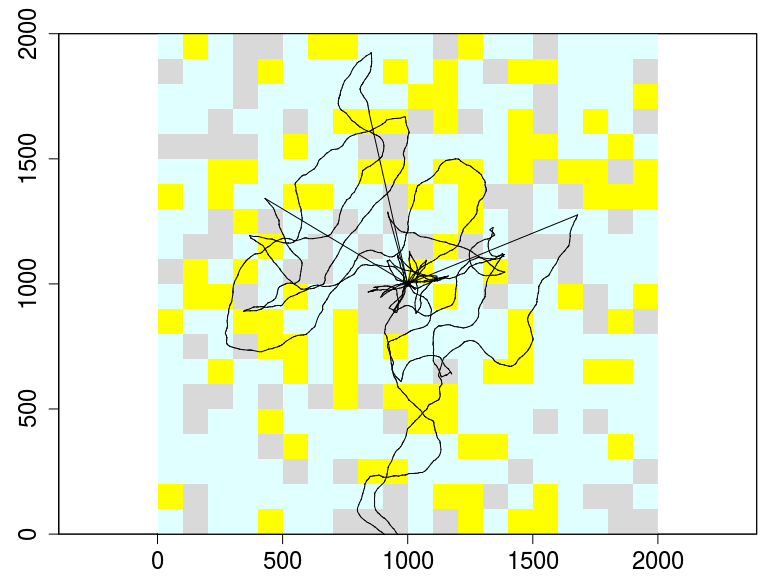} \qquad
	\includegraphics[width=.35 \textwidth]{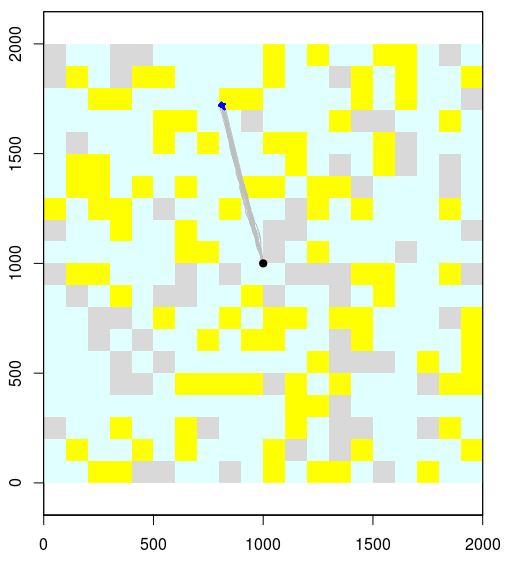}
	\caption{A set of exploratory trips (left) and foraging trips (right) on two 4 km$^2$ landscapes containing a mixture of crop and natural areas. The foraging trips include scouting movement to the foraging site, harvesting movement in the neighbourhood of the foraging site, then scouting movement back to the nest.  The harvesting movement is too small to be visible in the plot.  Blue areas are crop, yellow areas are natural, and gray areas contain no floral resources. The nest is located at the centre of the landscape.}\label{FigureCh3:FlightTracks}
\end{figure}

\begin{figure}
	\includegraphics[width=.5\textwidth]{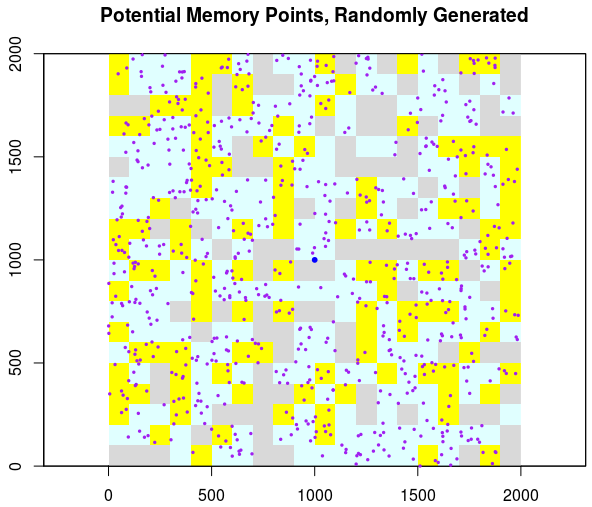}
	\includegraphics[width=.45\textwidth]{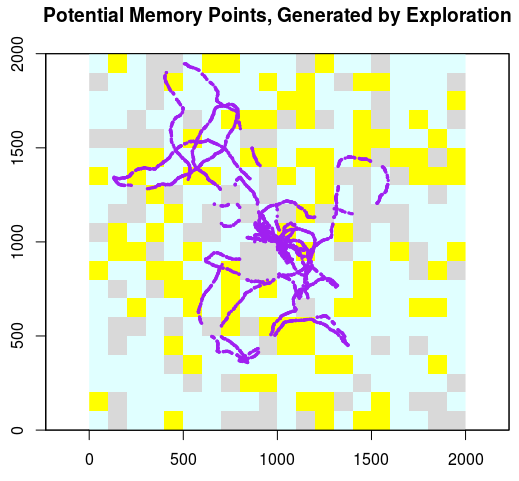}	\caption{The potential foraging sites (memory points) for a single bee generated randomly (left) or by exploration (right) on two 4 km$^2$ landscapes containing a mixture of crops and natural areas. Blue areas are crops, yellow areas are natural areas, and gray areas contain no floral resources. The nest is located at the centre of the landscape (blue dot), and the purple dots are potential points for sites of constancy.}\label{FigureCh3:MemoryCloud}
\end{figure}

The model tracks the movement behaviour of a single worker bee at a time. The first stage of the model is to create a list of potential foraging sites (the ``site list") using one of two site-generation methods, either ``random" or ``exploration."  Under random site-generation, the site list is created by randomly selecting points from throughout the domain, using a spatially uniform distribution, and excluding sites from areas containing no resources. Under exploration-based site-generation, the bee performs looping exploratory flights (Figure~\ref{FigureCh3:FlightTracks}, left) over the landscape, and  all resource sites encountered during these flights are included in the site list (Figure~\ref{FigureCh3:MemoryCloud}, right). In the earlier model \citep{MacQueen2021}, only exploratory flight behaviour was used; here, we compare results obtained using both random and exploratory site-generation. 

The second stage of the model is the selection of the site of foraging constancy (memory point) from among the set of points produced in the first stage. In the previous version of the model, the site of foraging constancy was selected randomly from within a single resource type, which was chosen based on nest needs and a simple preference for wild flowers. Here, we use four different algorithms for site-selection. These algorithms are detailed in Section~\ref{SectionCh3:Selection}. 

The third stage of the model is foraging, in which the bee makes seven foraging trips, representing a single day of foraging \citep{Woodgate2016}. These trips begin with a sequence of directed flight segments from the nest towards the site of foraging constancy. We refer to this behaviour as scouting\footnote{Note that we use the term ``scouting" differently than it is used in describing honey bee behaviour. Here, scouting refers only to the longer, directed flight segments used to get from the nest to the (known) foraging site or to move between foraging sites, and does not indicate searching behaviour.} \citep{tyson:2011}, i.e., directed movement toward the foraging site. Once the bee is near enough to her foraging site, she switches to harvesting \citep{tyson:2011}, i.e. moving randomly between flowers, collecting nectar at each flower that has not previously been visited by any bee. In this model, pollination (i.e. flower visiting) is limited to these harvesting segments. After completing the harvesting segment, the bee may then intersperse scouting flight segments with additional harvesting segments until she has collected her maximum load capacity (80 $\mu l$ of nectar). She then completes the foraging trip by returning directly to the nest. A sample set of foraging trips is shown in Figure~\ref{FigureCh3:FlightTracks} (right plot). 

Below, we describe the implementation of the four site-selection methods (Section~\ref{SectionCh3:Selection}) and the agricultural landscape (Section~\ref{SectionCh3:Landscapes}.  We then describe the methods we use for our sensitivity analysis (Section~\ref{SectionCh3:SensitivityAnalysis}), simulations (Section~\ref{SectionCh3:Simulations}), and statistical analyses (Section~\ref{SectionCh3:Stats}). 

\subsection{Selection of sites}
\label{SectionCh3:Selection}

Once the site list has been created (randomly or by exploration), the site of constancy is selected using one of four metrics: (1) none (site selected at random), (2) distance from the nest, (3) local density of wild flowers, or (4) net rate of energy return.  Note that while energy-based selection incorporates both distance from the nest and type-specific flower characteristics, including these mechanisms also as separate options allows for the possibility that they are significant to the bees for reasons other than energetic concerns. For example, sites could be selected based on distance from the nest due to predator avoidance and competition or resource depletion near the nest \citep{Goulson2010}. Site selection based on flower type (wild flower density) could result from other nutritional needs such as amino acid content in pollen or other required micronutrients \citep{Bobiwash2018,Vaudo2016}.

For  distance, wild flower density, and energy-based selection algorithms, we assume that the bees have incomplete knowledge about the potential sites, or are assessing the quality of a site imperfectly. To represent this, we use a probability distribution (Figure~\ref{FigureCh3:Optimization_All}) that assigns higher probability to better sites (according to the relevant selection method). Thus, the bees are more likely to select a ``good" site than a ``bad" one according to the relevant metric, but will not always select the ``best" site. 

Below, we discuss the relevance of each site-selection metric and its implementation in the model.

\subsubsection{Random}
\label{Section"RandomSelection}

Under the the ``random" metric, we assume that the forager has no preference for one site over another.  This method is unrealistic because it suggests that there is no behavioural component to the selection of the site of foraging constancy. However, random sampling is an approach often used in models for efficiency.  Here, we implement the random metric by selecting a site through random sampling from the site list, with all sites equally likely.

\subsubsection{Distance}
\label{SectionCh3:DistanceSelection}

Although bumble bees are central place foragers, they seem to be less likely to forage very close to the nest than slightly farther away, likely due to pressures from competition, resource depletion, and predation \citep{Dramstad1996, Osborne1999, Dukas1998, Goulson2010}. For this site-selection method, we therefore  hypothesize that bees are most likely to forage at some intermediate distance from the nest, avoiding competition but keeping flight costs low, following the beta distribution shown in Figure~\ref{FigureCh3:Optimization_All} (left). As the beta distribution ranges between 0 and 1, we normalize the distances so that the distance of the point farthest from the nest is set to 1. Selecting a site is then accomplished by randomly generating a number from the beta distribution, and selecting the site with the normalized distance closest to that number.

\subsubsection{Local wild flower density}
\label{SectionCh3:TypeSelection}

Bumblebees often show a clear preference for a particular resource, independent of its abundance. For example, bumble bees near blueberry crops, where blueberry flowers are by far the most abundant resource available, have been observed to collect disproportionately high amounts of non-blueberry pollen \citep{Bobiwash2018, Toshack2019}.  For this site-selection method, we therefore suppose that the bees prefer to forage in areas that are rich in wild flowers rather than crop flowers.  To determine the amount of wild flowers around each potential foraging site, we consider a 5 m by 5 m box around the point defining the site, and calculate the percent of the box that is occupied by wild flowers. We then sort the sites by percentage of wild flowers into 10 bins of equal size (i.e. bin 1 = $[0\%,10\%)$, bin 2 = $[10\%,20\%)$, and so on). To select a point, we first select a bin using a discrete probability distribution (Figure~\ref{FigureCh3:Optimization_All} (right)), and then randomly select one of the points in that bin. The probability distribution for bin selection gives greater probability to bins with a higher percentage of wild flowers  (Figure~\ref{FigureCh3:Optimization_All}). 
 For simplicity, we do not also consider the depletion of resources, so two locations with equal amounts of wild flower habitat, only one of which has been visited by other bees, are treated equally.

\subsubsection{Energy}
\label{SectionCh3:EnergySelection}

Several studies have suggested that bee foraging decisions are strongly influenced by energetic considerations \citep{Cartar1991, Cresswell2000, Charlton2010}. In our implementation of this site-selection method, we assume that bumble bees are optimizing their net rate of energy return, ignoring all other considerations. We follow the model proposed by \citet{Cresswell2000}. That is, 
\begin{equation}
\label{Equation:EnergyRate}
\Psi = \frac{C\Lambda - \frac{D}{S}(M_S) - PM_H}{\frac{D}{S}+P},
\end{equation}
where $P = \frac{HC}{R}$ is the total harvesting time ($s$), $H$ is the time per flower ($s$), which includes both handling time and time flying between flowers, $C$ is how much resource the bee can carry ($\mu l$ nectar), $R$ is the amount of resource collected from a single flower ($\mu l$ nectar), $\Lambda$ is the metabolic energy per $\mu l$ of nectar ($J$), $D$ is the round-trip distance to and from the site ($m$), $S$ is the flight speed while travelling to the location ($m/s$), $M_S$ is the rate of energy expenditure while flying ($J/s$), and $M_H$ is the rate of energy expenditure while harvesting ($J/s$). Default values for each of these parameters are provided in Table~\ref{TableCh3:EnergyParameters}. Details on the sources for and calculation of these values are provided in the Supporting Information. Note that this model assumes that a foraging trip ends only when the bee has a completely full honey crop. 

\begin{figure}
	\includegraphics[width=.45\textwidth]{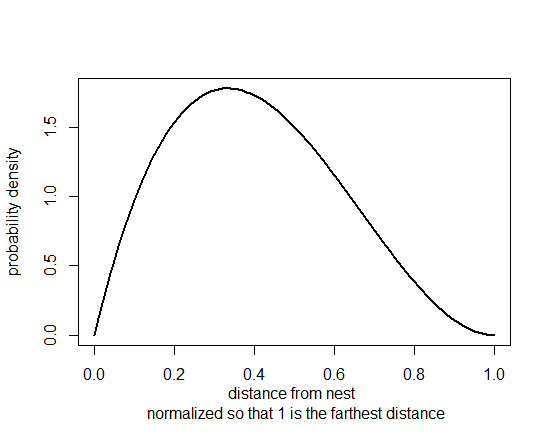} \hfill 
	\includegraphics[width=.45\textwidth]{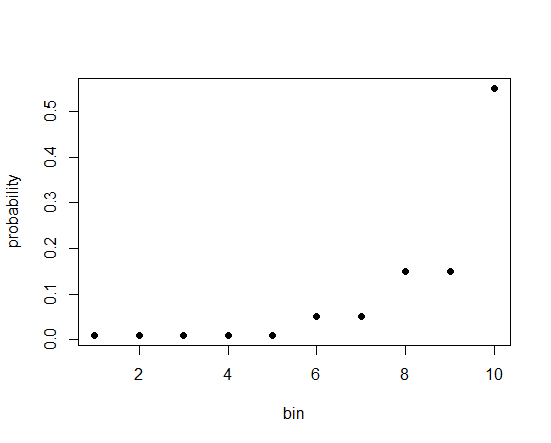}
	\caption{The continuous (left) and discrete (right) probability distributions used in selecting the site of constancy.  Selection may be based on the distance to the site (left), or on binning of the local wild flower density or the net rate of energy return at the site (right). The probability distribution on the left is a $\beta$-distributions with parameters  $\alpha=2$ and $\beta=3$ .  For sorting of the local wild flower density or energy return values, the points in the site list are first sorted into 10 equal sized bins of increasing energy or density, and the probability distribution is used to select the bin, before the site of constancy is selected randomly from the bin. (See Section~\ref{Appendix:discrete} for reasoning behind using discrete distributions for wildflower density and energy.)}\label{FigureCh3:Optimization_All}
\end{figure}

As the nectar per flower, handling time, and energy per $\mu l$ of nectar generally differ for blueberry flowers and wild flowers, we calculate the net rate of energy return as an average of the net rate of energy return when harvesting solely on blueberry flowers ($\Psi_{blb}$) and the net rate of energy return when harvesting solely on wild flowers ($\Psi_{wf}$), weighted by the fraction of blueberry flowers ($\alpha_{blb}$) and wild flowers ($\alpha_{wf}$) present at the site. That is,
\begin{equation}
\Psi_{site} = \dfrac{\alpha_{blb}\Psi_{blb} + \alpha_{wf}\Psi_{wf}}{2}.
\end{equation}
We then select the site of constancy in the same way as in the local wild flower density scenario (Section~\ref{SectionCh3:TypeSelection}), by first sorting the points according net rate of energy return into 10 bins of equal size (such that bin 1 has the lowest $10\%$ of the points and bin 10 has the highest $10\%$ of the points), then selecting the bin using the probability distribution shown in Figure~\ref{FigureCh3:Optimization_All}, and finally selecting the foraging site randomly from within the chosen bin.
\begin{table}
	\centering
	\caption{The parameters for the net rate of energy gain from a foraging site (Equation~\eqref{Equation:EnergyRate}).}\label{TableCh3:EnergyParameters}
	\begin{subtable}{\textwidth}
		\centering
		\caption{Flower parameters. Note that we use parameters for nectar throughout, as these are the most readily available in the literature and are required in Equation~\ref{Equation:EnergyRate}. However in discussion of the model, we do not distinguish between nectar and pollen, and simply refer to ``resource" to indicate this.}
       \rowcolors{2}{white}{gray!15!}
       \begin{tabularx}{\linewidth}{
   >{\raggedright\arraybackslash\setlength{\hsize}{1\hsize}}X
 | >{\raggedright\arraybackslash\setlength{\hsize}{0.5\hsize}}X
 | >{\raggedright\arraybackslash\setlength{\hsize}{0.7\hsize}}X  
 | >{\raggedright\arraybackslash\setlength{\hsize}{0.5\hsize}}X  
 | >{\raggedright\arraybackslash\setlength{\hsize}{2.3\hsize}}X  }
			Parameter & Symbol  & Wild flower Value & Blueberry Value & References \\ \hline
			nectar per flower  & $R$ ($\mu l$)& 0.8 & 0.8 & \cite{Hodges1981,Cresswell2000} \\ 
			handling time  & $H$ ($s$)& 5 & 7.67 &  \cite{Cresswell2000,RodriguezSaona2011} \\ 
			energy per $\mu l$ nectar  & $\Lambda$ ($J$)& 3.99 & 8.746 & \cite{Brewer1969,Cresswell2000,Becher2018} \\ 
		\end{tabularx}
	\end{subtable}
	
	\vspace{.5cm}
	
	\begin{subtable}{\textwidth}
		\centering
		\caption{Bee parameters}
       \rowcolors{2}{white}{gray!15!}
       \begin{tabularx}{\linewidth}{
   >{\raggedright\arraybackslash\setlength{\hsize}{1.5\hsize}}X
 | >{\raggedright\arraybackslash\setlength{\hsize}{0.5\hsize}}X
 | >{\raggedright\arraybackslash\setlength{\hsize}{0.5\hsize}}X  
 | >{\raggedright\arraybackslash\setlength{\hsize}{1.5\hsize}}X   }
			Parameter & Symbol  & Value & References \\ \hline
			amount bee can carry & $C$ ($\mu l$) & 60  & \cite{Allen1978} \\ 
			flight speed & $S$ ($m/s$)& 4.5 & \cite{Dudley1990} \\ 
			rate of energy expenditure while flying & $M_S$ ($J/g/s$) & 0.33 & \cite{Heinrich1975,Cartar1991} \\ 
			rate of energy expenditure while harvesting & $M_H$ ($J/g/s$) & 0.235 & \cite{Heinrich1975,Cartar1991} \\ 
		\end{tabularx}
	\end{subtable}
\end{table}

\subsection{Landscapes}
\label{SectionCh3:Landscapes}

We parametrize the model for blueberry crops and generic  ``wild flowers," that do not represent any single wild flower species, but rather a composite of several species. The handling time comes from \cite{Cresswell2000}, who use an average of six species, and the energy per $\mu$l nectar is calculated based on the average sugar concentration of the wildflower species listed in \cite{Becher2018}.  
The landscape is 2 km by 2 km, and made up of 100 m by 100 m cells. The resource type of each cell is randomly determined, with probabilities $p_b = 1/3$ of being blueberry, $p_w = 1/3$ of being wild flower, and $p_e = 1/3$ of being empty (having no floral resources). A different random landscape is generated for each model simulation

\subsection{Sensitivity Analysis}
\label{SectionCh3:SensitivityAnalysis}
We ran a full sensitivity analysis on the model following the procedure outlined by \cite{Prowse2016}. Details, including a list of the parameters tested, is in the Supporting Information.

\subsection{Simulations}
\label{SectionCh3:Simulations}

For each combination of site-generation and site-selection method, we ran 20 simulations. In each simulation, a single nest with 50 worker bees is placed at the centre of the landscape, and the bees forage one at a time. The bees do not interact directly in any way, but can interact indirectly via nectar depletion on the landscape, as any flowers visited by a bee are considered depleted for the rest of the simulation, and will not be visited by any other bees. For each simulation, we record the following data: (1) the location of each foraging site, (2) the average composition of pollen loads brought back to the nest by each bee, (3) the locations of all flower visits, and (4) the amount of time spent in each different movement mode (harvesting, scouting, returning to the nest).

\subsection{Statistical Analyses}
\label{SectionCh3:Stats}

Using the data recorded for each simulation (see Section~\ref{SectionCh3:Simulations}), we calculate the total number of blueberry flowers visited, the percent of foraging sites located in blueberry flowers, the percent of blueberry in the resource loads, the average percent of time spent harvesting (visiting flowers), and the average total amount of time spent foraging.
\noindent To measure the relative importance of site-generation method and site-selection method, we take an approach similar to that of \citet{Corell2012} and \citet{Everaars2018}, and report the amount of variability explained by each factor or combination of factors, determined by performing a two factor analysis of variance (ANOVA) on each response. To assess the relative impact of the site-generation method and site-selection, we perform a graphical analysis of variance (GANOVA) for each metric, and visually compare the relative difference in outcome across treatments. These GANOVA plots display the amount of variability in the individual measurements as a histogram of residuals and compare it to the amount of variability in the means of each treatment group, displayed as symbols plotted across the $x$-axis of the histogram. A wide spread of the means relative to the residuals indicates evidence of differences in the means. Additionally, the direction of the differences can be observed from the values of the means, and any groupings of treatments that are not different from one another can be observed as clusters of means. (See \cite{MacQueen2021,Braun2012,Box2005} for further explanations of the GANOVA method.)

\section{Results}
\label{SectionCh3:Results}

\begin{figure}
	\centering
	\begin{subfigure}[b]{.47\textwidth}
		\includegraphics[width=\textwidth]{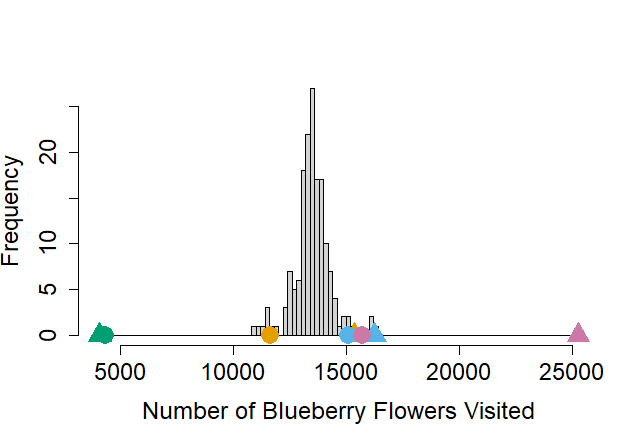}
		\caption{}\label{FigureCh3:GANOVA_Flowers}
	\end{subfigure}
	\begin{subfigure}[b]{.47\textwidth}
		\includegraphics[width=\textwidth]{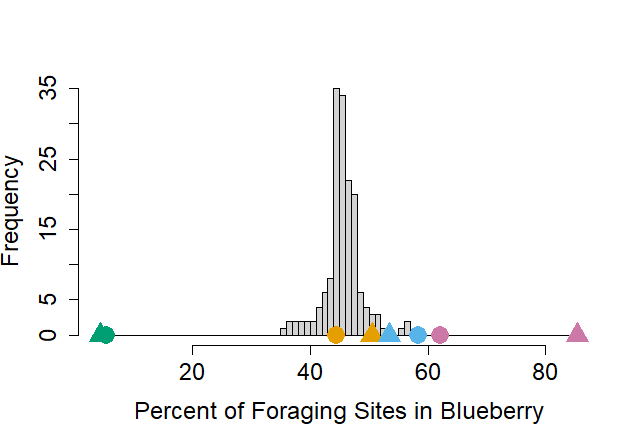}
		\caption{}\label{FigureCh3:GANOVA_Sites}
	\end{subfigure}
	
	\begin{subfigure}[c]{.47\textwidth}
		\includegraphics[width=\textwidth]{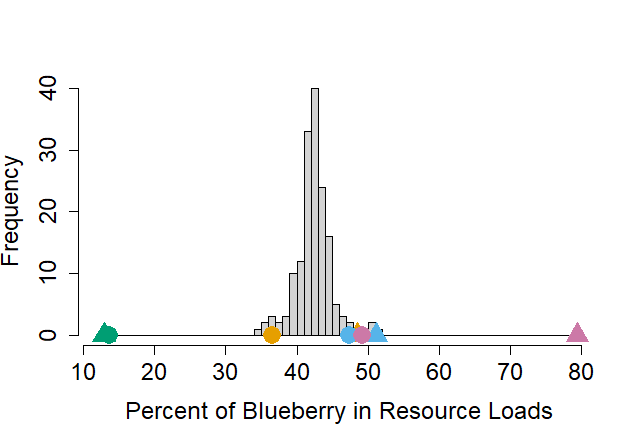}
		\caption{}\label{FigureCh3:GANOVA_Loads}
	\end{subfigure}
	\begin{subfigure}[c]{.47\textwidth}
		\includegraphics[width=\textwidth]{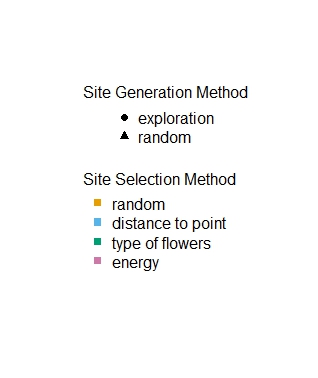}
	\end{subfigure}
	
	\begin{subfigure}[b]{.47\textwidth}
		\includegraphics[width=\textwidth]{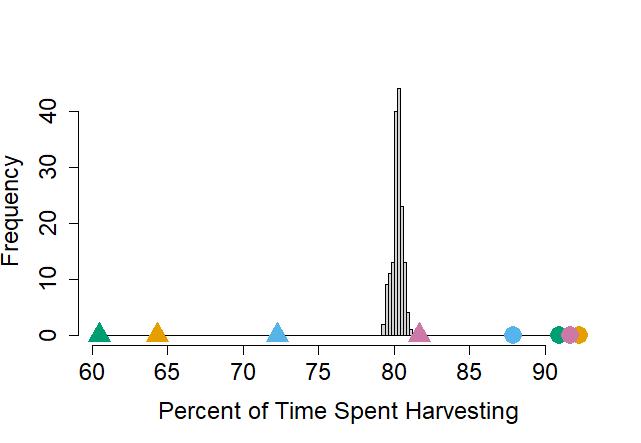}
		\caption{}\label{FigureCh3:GANOVA_Harvesting}
	\end{subfigure}
	\begin{subfigure}[b]{.47\textwidth}
		\includegraphics[width=\textwidth]{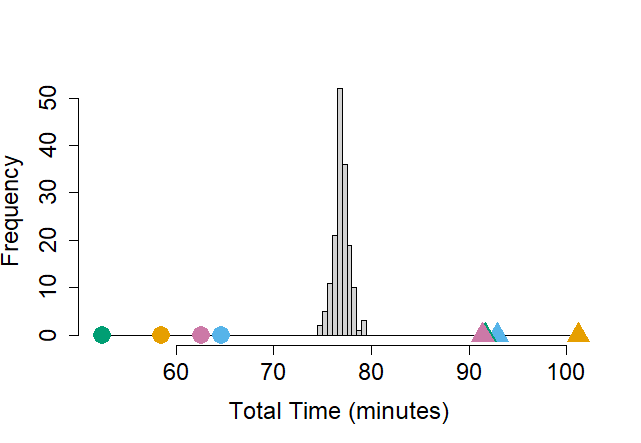}
		\caption{}\label{FigureCh3:GANOVA_TotalTime}
	\end{subfigure}
	\caption{GANOVA plots for each different response. The symbols on the x-axis represent the means of each treatment at the correct value. This histogram shows the spread of the entire data set with the residuals (treatment mean - individual measurement). Any two treatment means that are farther apart than the spread of the data can be considered significantly different \citep{MacQueen2021,Braun2012,Box2005}.}
    \label{FigureCh3:GANOVAs}
\end{figure}

Figure~\ref{FigureCh3:GANOVAs} shows the mean values of each model response variable, for each treatment. Table~\ref{TableCh3:Variability} lists the percent of variability in each response variable due to site-generation method, site-selection method, and the combination of the two. Any entry in Table~\ref{TableCh3:Variability} that appears as a measurable effect in Figure~\ref{FigureCh3:GANOVAs} is marked in bold.  The site-generation method and the site-selection method together explain approximately 76-98\% of the variability in the model response, depending on the response.

For the number of blueberry flowers visited, the percent of foraging sites in blueberry, and the percent of blueberry in the resource loads, the site-selection method is significant, while the site-generation method is not. From Figure~\ref{FigureCh3:GANOVAs}, subplots (\subref{FigureCh3:GANOVA_Flowers})-(\subref{FigureCh3:GANOVA_Loads}), we see that site-selection based on energy leads to the highest mean value for all three responses, selection based on the type of flowers leads to the lowest mean values, and selection based on distance to the foraging site or random selection leads to mean values that fall in the middle. 

For the percent of time spent harvesting, the site-generation method is highly significant, and the site-selection method is somewhat signifcant, as is the interaction between site-generation and site-selection. For site-generation based on exploration, each selection method leads to mean values that differ substantially from the others (Figure~\ref{FigureCh3:GANOVAs}(\subref{FigureCh3:GANOVA_Harvesting})).  Specifically, the ordering of site-selection methods for the mean time spent harvesting is type of flowers, random, distance from the nest, and net rate of energy return.  Site-generation based on exploration yields higher values for the mean percent of time spent harvesting than random site-generation, regardless of the site-selection method used. 

For the total trip time, which includes the harvesting and scouting portions of each foraging bout, only the site generation method is significant, with random site generation leading to longer foraging trips. For site-selection based on exploration, the ordering of mean values (Figure~\ref{FigureCh3:GANOVAs}(\subref{FigureCh3:GANOVA_TotalTime})) is type of flowers, random, energy, distance from the nest, but the distance between neighbouring means is not significant.


\subsection{Sensitivity analysis results}

The relative influence of each parameter for each response variable in each of the eight combinations of site generation / site election can be found in S03. This file contains eight spreadsheets, one for each combination, plus a summary listing the parameters with at least 10\% relative influence for each combination. 

No clear pattern emerges, but only a subset of 15 parameters have a relative influence greater than 10\%  in any case. There are 40 combinations of site generation method, site selection method, and model output (metric). Of these, the parameter $R_w$ is of influence 19 times; $s$, {\em too long}, and $\mu$ 12 times each; $R_b$ 10 times; $H_b$ 7 times; $\gamma$ and $D_i$ 6 times; $rd_w$, $rd_b$, and $e_i$ 2 times; and $M_H$, $M_S$, {\em too long e}, and $\Lambda_B$ 1 time. The parameter $\mu$ is mostly of influence in the cases with random site generation, and of course the parameters pertaining to exploratory behaviour ($D_i$ and $e_i$) are only of significance when exploratory site generation is used. 

We describe here the effects of the most significant parameter, $R_w$ and the effects of the exploratory parameters, which were not previously tested \citep{MacQueen2021}.

The parameter $D_i$ controls how much bigger successive exploratory loops get (irrespective of the starting size). The parameter $e_i$ controls how much longer the exploration trips get (irrespective of the starting length). Across all site selection methods, as $D_i$ increases, the percent of time spent harvesting, the total number of sites in blueberry, and the percent of blueberry in the resource loads all decrease and as $e_i$ increases, the total number of sites in blueberry decreases. 

The parameter $R_w$ is the amount of nectar collected per wildflower. $R_w$ is never influential (at greater than 10\%) for the percent of time spent harvesting. This is because the quantity of nectar collected does not, in our model, affect the behaviour of the bee (e.g. the handling time). As $R_w$ increases, the total trip time and the number of blueberry flower visits decrease, because the bee is getting more nectar per flower and therefore getting a full resource load faster.

\begin{table}
\caption{The percent of the variability calculated from the analysis of variance for each response variable (column). Any effects that appeared measurably in Figure~\ref{FigureCh3:GANOVAs} are marked in bold. }\label{TableCh3:Variability}
       \rowcolors{2}{white}{gray!15!}
       \begin{tabularx}{\linewidth}{
   >{\raggedright\arraybackslash\setlength{\hsize}{1.5\hsize}}X
 | >{\raggedright\arraybackslash\setlength{\hsize}{0.9\hsize}}X
 | >{\raggedright\arraybackslash\setlength{\hsize}{0.9\hsize}}X  
 | >{\raggedright\arraybackslash\setlength{\hsize}{0.9\hsize}}X   
 | >{\raggedright\arraybackslash\setlength{\hsize}{0.9\hsize}}X   
 | >{\raggedright\arraybackslash\setlength{\hsize}{0.9\hsize}}X   
 }
Source of variation & Blueberry flowers visited & Percent of foraging sites in blueberry & Percent of blueberry in resource loads & Percent of time harvesting & Total time \\ \hline
Site generation method & 5.77 & 0.97 & 5.96 & {\bf 74.91} & {\bf 90.53} \\
Site selection method & {\bf 64.09} & {\bf 73.27} & {\bf 63.97} & {\bf 11.22} & 2.66 \\
Generation : selection & 6.47 & 3.32 & 6.50 & {\bf 12.19} & 3.01 \\
Residual & 23.68 & 22.43 & 23.57 & 1.68 & 3.80 \\
\end{tabularx}
\end{table}

\section{Discussion}
\label{SectionCh3:Conclusion}

{\rct How foragers use the landscape to acquire resources is an important and well-studied topic \citep{abrahms:2021, dwinnell:2019, wam:2018, covich:1976}, especially in the context of climate change \citep{knowlton:2010}.  Some organisms demonstrate `` site constancy", i.e., foraging behaviour that is characterised by repeated visits to a particular location \citep{BergerTal2015, Baylis2015, Chilvers2008, Crist1991, Thomson1997}.
Here we are interested in site-constant foragers that select a foraging site after first engaging in a separate exploration phase (see, e.g., \citet{popp:2023, nowak:2023, osborne:2013, carter:2020, mandal:2019}).  In these cases, the foraging site is likely selected from among those locations visited during exploration.  From a modelling perspective, we are then led to ask if if is necessary to model this exploration phase explicitly, or if it is sufficient to simply select a foraging site at random from among logical options.  In other words, we ask if the details of exploratory movement (site-generation and subsequent site-selection) affect the outcomes (e.g., fitness, landscape use) of site-constant foraging movement?

We take the bumble bee as our model organism, as the exploration and site-constant foraging phases of movement have been well documented for this organism \citep{woodgate:2016, lihoreau:2012}.  In this context, landscape use is represented by pollination services \citep{woodard:2017}.  Many economically important crops are heavily reliant on bumble bee pollination services \citep{Garibaldi2013}, and models are useful tools in predicting both the survival of bee populations and the pollination services they provide \citep{Everaars2014, MacQueen2021, carturan:2023, Becher2018,twiston-davies:2021, Lonsdorf2009}.  One such crop is highbush blueberry \citep{Blaauw2014}, and so, for concreteness, this is the crop in our simulations.
}

 Our results show that the site-selection method for the site of foraging constancy strongly affects  the  pollination services (number of blueberry flower visits and percent of foraging sites in blueberry) and moderately affects the bee behavioural budget (percent of time spent harvesting) predicted by the model. Conversely, the site-generation method strongly affects the bee behavioural budget but not the pollination services. Bee fitness is more mixed, with the site-selection method strongly affecting the percent of blueberry in the bee resource loads, and the-site generation method strongly affecting the total foraging trip time. It is therefore critical that both elements of bumble bee behaviour be understood and modelled correctly. 
	
When selecting a foraging site based on energy, the bees are more likely to select foraging sites in blueberry, probably because the higher energy content of blueberry nectar outweighs the energy demand of its longer handling time. This choice also increases the visits to blueberry flowers, and therefore also the amount of blueberry nectar the bees collect. Site selection based on the local density of wild flowers leads to the lowest crop pollination services and the lowest amount of blueberry resource collected, because of bee preference for wild flowers in this system. Random selection and distance-based site-selection fall in the middle and are not significantly different from one another. There is no reason for site-selection based only on distance to the nest to have any bias for either type of resource in our model landscapes, and accordingly selection based on these methods both result in about 50\% of foraging sites in blueberry and 50\% blueberry in the resource loads collected by the bees. 

Exploratory site-generation behaviour tends to generate more potential sites closer to the nest than the random site-generation does. Therefore, in the treatments with exploratory site-generation, the bees are more likely to pick a site closer to the nest, and, on average, travel a shorter distance to their foraging site, leading to shorter trips overall.  The site selection method has minimal impact on total trip time in comparison to the site generation method. The shorter distance travelled in treatments with exploratory site generation also means that the bee spends proportionally less time scouting and returning, and more time harvesting. The site selection method has minimal effect on the percent of time spent harvesting when exploratory site generation method is used, but does matter for random site generation. Selection based on energy leads to a greater number of blueberry flower visits, and since blueberry flowers have a longer handling time, the proportion of harvesting time is greater on average. Conversely, selecting based on type of flowers leads to more wildflower visits, and a smaller percent of harvesting time on average. 

The crop we model here is an economically valuable crop \citep{tootelian:2020, statscan:2022} that benefits from bumble bee pollination, but that is also nutritionally deficient for bees \citep{Roger2017}.  So the bees are attracted to the high energy density and abundance of the crop flowers, but preferentially seek out wild flower patches to obtain a balanced protein diet.  Since many popular highbush blueberry varietals are better served by bumble bees than managed honey bees \citep{Javorek2002}, our results are particularly relevant to pollination services modelling for this industry.  While we focussed on a single type of crop in our study, our work shows that site-generation and site-selection methods can significantly affect crop pollination services in this system, and so are likely important in other systems as well.

\subsection{Implications for field studies}
	

Our model results could be compared to field data to help determine the factors that affect bee decision making.  To illustrate this process, 
we look at the studies carried out by \citet{Bobiwash2018} and \citet{Toshack2019}. In both of these studies, it is observed that bumble bees in areas where a high proportion of the landscape is planted with blueberry forage proportionally less on blueberry flowers than on the other available flowers.  
Our model predicts the highest numbers of visits to blueberry flowers under energy based site-selection, suggesting that if energy were the main factor driving the bee foraging choices, the bees in the \citet{Bobiwash2018} and \citet{Toshack2019} studies should have been visiting the blueberry crops more than the other available flowers.  Since they were not, we conclude that energy is probably not the only factor driving bee selection of foraging sites in this situation. In fact, in our model, random site-selection, selection based on distance, and selection based on wild flower density all lead to resource loads of similar composition to those observed in \citet{Bobiwash2018}. This pattern suggests that bumble bees in blueberry crop systems may be using some combination of distance from the nest and local wild flower density to select foraging sites. This result stands in contrast to other research showing that bees forage based on energy (e.g. \citet{Cartar1991,Charlton2010}), although of course it relies on the assumptions we have made regarding the energetic contents of blueberry and wild flower resources. 
A difference between our work and earlier studies \citep{Cartar1991, Charlton2010} is the focus on blueberry crops, whose pollen has low crude protein content and lacks seeral of the amino acids that are essential for bees \citep{Toshack2019, Somerville2001, Somerville2006, Dufour2020}. We hypothesize that results similar to ours would be obtained for other crop systems, such as almond, that are also of poor nutritional quality for bees.

 In addition to comparing the distribution of resource types gathered by the bees, we can also use other measures of bee fitness to suggest how real bees might select their foraging sites.  To make these predictions, we assume that the bees will use the site-selection method that leads to the highest fitness.  If we suppose fitness is measured in percent of time spent harvesting, such that a higher percent of time harvesting indicates higher fitness, then exploratory site-generation  with site-selection based on energy, wild flower density, or random selection are the most likely methods. Overall, when exploratory site-generation is used rather than random site-generation, all four site-selection methods lead to a higher percent of time spent harvesting.   If we suppose instead that the highest fitness is given by the lowest total foraging trip time, then exploratory site-generation with any of the site-selection methods is most likely. Exploratory site-generation leads to a higher percent of time spent harvesting as well as a shorter total trip time for all site-selection methods, so it seems likely that real bees perform site-generation in this way.

\subsection{Implications for modelling}

{\rct 
Our results suggest that
the manner in which site-generation and site-selection are implemented can have a strong effect on model outcomes, though the two behaviours aren't necessarily equally important. Indeed, the relative importance of correctly modelling site-generation and site-selection depends on the output metrics of the model.  For example, if the goal is to measure visits to a particular type of resource patch,  i.e., pollination services, then an accurate representation of the site-selection method is most important.  If, on the other hand, the goal is to measure the behavioural energy budget of the forager, then the an accurate representation of the site-generation method is most important.  
}


{\rct When creating a behavioural, mechanistic model, it is impossible, and even undesirable, to model all behaviours exactly as they occur in real life. Instead, the research question can help determine which behaviours should be included, and which are not essential to answering the question \citep{Railsback2012}.  
Most modelling studies of site-constant foraging do not explicitly include a separate exploration phase as part of the identification of a foraging location.  Our work demonstrates that the omission of site-generation and site-selection behaviours is potentially creating biases in predictions for certain types of landscape use and can dramatically alter predicted foraging outcomes.  
}  


\section{Acknowledgements} 
Funding for this work was provided by the Natural Sciences and Engineering Research Council of Canada (RCT) through a Strategic Project Grant (506922-17) and a Discovery Grant (RGPIN-2016-05277).  In addition, SAM received a Mitacs Globalink Research Award (BC/ISED).  The authors also gratefully acknowledge the support of the UBC Okanagan Institute for Biodiversity Resilience and Ecosystem Services.

\section{Author Contributions}
This work is part of SAM's PhD thesis.  SAM designed the study, wrote the code, did the analysis, and wrote the first draft of the paper.  RCT and WJB were SAM's thesis supervisor and co-supervisor, respectively, and were actively involved at a supervisory level in the study design and analysis of the results.  SAM, CFH, and RCT contributed to the editing process.

\section{Competing Interests}
The authors declare that they have no competing interests to declare.

\bibliography{SiteConstancy_TheorEcol}

\appendix

\section{Discrete Probability Distributions for wildflower density and energy based site selection}
\label{Appendix:discrete}

Suppose we are selecting a foraging site based on the local wildflower density, and let X be the amount of wildflower resource in the area around a single potential foraging site. Using a continuous distribution, $P(X=a)=0$ for any single value of $a\geq0$, but $P(a<X<b)=p$, where $p>0$. The two probability distributions we use have increasing probability with increasing wildflower density. Therefore, a site with wildflower density between 0.9 and 0.99 is more likely to be selected than either a point with a lower wildflower density, or a point with entirely wildflower resource. This, then, leads to over-selecting points that lie near the edges of a wildflower field. Using instead a discrete distribution, with sites sorted into bins, we can give equal probability to all points that have wildflower densities of 0.9 to 1, and avoid over-selecting the edges of the fields. 

A similar argument applies to the energy distributions, noting that a point with 100\% blueberry flowers has a higher net rate of energy return than a point with 100\% wildflowers. Thus, in the continuous case, this leads to an over-selection of points on the edges of blueberry fields (which have, say, 90-99\% of the maximum possible energy).

\end{document}